\documentclass[twocolumn]{aastex63}

\usepackage{xcolor}
\usepackage{CJK}

\newcommand\msun{$\rm M_{\odot}$}

\received{September 1, XXXX}
\revised{October 1, XXXX}
\accepted{November 1, XXXX}

\submitjournal{XXX}

\shorttitle{SN~2016dsg}
\shortauthors{Dong et al.}
\graphicspath{{./}{figures/}}

\begin{document}

\title{SN~2016dsg: A Thermonuclear Explosion Involving A Thick Helium Shell}
\begin{CJK*}{UTF8}{gbsn}

\correspondingauthor{Yize Dong}
\email{yizdong@ucdavis.edu}

\author[0000-0002-7937-6371]{Yize Dong (董一泽)}
\affiliation{Department of Physics and Astronomy, University of California, 1 Shields Avenue, Davis, CA 95616-5270, USA}
\author[0000-0001-8818-0795]{Stefano Valenti}
\affiliation{Department of Physics and Astronomy, University of California, 1 Shields Avenue, Davis, CA 95616-5270, USA}

\author[0000-0002-1633-6495]{Abigail Polin}
\affiliation{The Observatories of the Carnegie Institution for Science, 813 Santa Barbara St., Pasadena, CA 91101, USA}
\affiliation{TAPIR, Walter Burke Institute for Theoretical Physics, 350-17, Caltech, Pasadena, CA 91125, USA}

\author[0000-0002-7537-6921]{Aoife Boyle}
\affiliation{AIM, CEA, CNRS, Universit\'e Paris-Saclay, Universit\'e Paris Diderot, Sorbonne Paris Cit\'e, F-91191 Gif-sur-Yvette, France}

\author[0000-0003-2024-2819]{Andreas Fl\"ors}
\affiliation{GSI Helmholtzzentrum f\"ur Schwerionenforschung, Planckstra{\ss}e 1, 64291 Darmstadt, Germany}

\author[0000-0002-7941-5692]{Christian Vogl}
\affiliation{Max-Planck-Institut f\"ur Astrophysik, Karl-Schwarzschild-Str. 1, 85748 Garching, Germany}
\affiliation{Exzellenzcluster ORIGINS, Boltzmannstr. 2, D-85748 Garching, Germany}

\author{Wolfgang E.\ Kerzendorf}
\affiliation{Department of Physics and Astronomy, Michigan State University, East Lansing, MI 48824, USA}

\author[0000-0003-4102-380X]{David J.\ Sand}
\affiliation{Steward Observatory, University of Arizona, 933 North Cherry Avenue, Rm. N204, Tucson, AZ 85721-0065, USA}

\author[0000-0001-8738-6011]{Saurabh W.\ Jha}
\affiliation{Department of Physics and Astronomy, Rutgers, the State University of New Jersey, 136 Frelinghuysen Road, Piscataway, NJ 08854, USA}

\author[0000-0002-9658-6151]{\L{}ukasz Wyrzykowski}
\affiliation{Astronomical Observatory, University of Warsaw, AI. Ujazdowskie 4, 00-478 Warszawa, Poland}

\author[0000-0002-4924-444X]{K.\ Azalee Bostroem}
\altaffiliation{DiRAC Fellow}
\affiliation{Department of Astronomy, University of Washington, 3910 15th Avenue NE, Seattle, WA 98195-0002, USA}

\author[0000-0002-0744-0047]{Jeniveve Pearson}
\affiliation{Steward Observatory, University of Arizona, 933 North Cherry Avenue, Rm. N204, Tucson, AZ 85721-0065, USA}

\author{Curtis McCully}
\affiliation{Department of Physics, University of California, Santa Barbara, CA 93106-9530}
\affiliation{Las Cumbres Observatory, 6740 Cortona Drive, Suite 102, Goleta, CA 93117-5575, USA}

\author[0000-0003-0123-0062]{Jennifer E.\ Andrews}
\affiliation{Gemini Observatory, 670 North A`ohoku Place, Hilo, HI 96720-2700, USA}


\author[0000-0002-3256-0016]{Stefano Benetti}
\affiliation{INAF, Osservatorio Astronomico di Padova, Vicolo dell'Osservatorio 5, I-35122 Padova, Italy}

\author[0000-0002-9388-2932]{St\'ephane Blondin}
\affiliation{Aix Marseille Univ, CNRS, CNES, LAM, Marseille, France}

\author[0000-0002-1296-6887]{L. Galbany}
\affiliation{Institute of Space Sciences (ICE, CSIC), Campus UAB, Carrer de Can Magrans, s/n, E-08193 Barcelona, Spain}
\affiliation{Institut d'Estudis Espacials de Catalunya (IEEC), E-08034 Barcelona, Spain}

\author[0000-0002-1650-1518]{Mariusz Gromadzki}
\affiliation{Astronomical Observatory, University of Warsaw, AI. Ujazdowskie 4, 00-478 Warszawa, Poland}

\author[0000-0002-0832-2974]{Griffin Hosseinzadeh}
\affiliation{Steward Observatory, University of Arizona, 933 North Cherry Avenue, Rm. N204, Tucson, AZ 85721-0065, USA}

\author[0000-0003-4253-656X]{D.\ Andrew Howell}
\affiliation{Department of Physics, University of California, Santa Barbara, CA 93106-9530}
\affiliation{Las Cumbres Observatory, 6740 Cortona Drive, Suite 102, Goleta, CA 93117-5575, USA}

\author[0000-0002-3968-4409]{Cosimo Inserra}
\affiliation{Cardiff Hub for Astrophysics Research and Technology, School of Physics \& Astronomy, Cardiff University, Queens Buildings, The Parade, Cardiff, CF24 3AA, UK}

\author[0000-0001-5754-4007]{Jacob E.\ Jencson}
\affiliation{Steward Observatory, University of Arizona, 933 North Cherry Avenue, Rm. N204, Tucson, AZ 85721-0065, USA}

\author[0000-0001-9589-3793]{Michael Lundquist}
\affiliation{W.~M.~Keck Observatory, 65-1120 M\=amalahoa Highway, Kamuela, HI 96743-8431, USA}

\author[0000-0002-3464-0642]{J. D. Lyman}
\affiliation{Department of Physics, University of Warwick, Coventry CV4 7AL, UK}

\author[0000-0002-0629-8931]{Mark Magee}
\affiliation{Institute of Cosmology and Gravitation, University of Portsmouth, Burnaby Road, Portsmouth, PO1 3FX, UK}

\author[0000-0002-9770-3508]{Kate Maguire}
\affiliation{School of Physics, Trinity College Dublin, The University of Dublin, Dublin 2, Ireland}

\author[0000-0002-7015-3446]{Nicolas Meza}
\affiliation{Department of Physics and Astronomy, University of California, 1 Shields Avenue, Davis, CA 95616-5270, USA}

\author[0000-0003-4524-6883]{Shubham Srivastav}
\affiliation{Astrophysics Research Centre, School of Mathematics and Physics, Queen's University Belfast, Belfast BT7 1NN, UK}

\author[0000-0002-4265-1958]{Stefan Taubenberger}
\affiliation{Max-Planck-Institut f\"ur Astrophysik, Karl-Schwarzschild-Str. 1, 85748 Garching, Germany}

\author[0000-0001-9834-3439]{J.~H.~Terwel}
\affiliation{School of Physics, Trinity College Dublin, The University of Dublin, Dublin 2, Ireland}

\author[0000-0003-2732-4956]{Samuel Wyatt}
\affiliation{Steward Observatory, University of Arizona, 933 North Cherry Avenue, Rm. N204, Tucson, AZ 85721-0065, USA}

\author[0000-0002-1229-2499]{D. R. Young}
\affiliation{Astrophysics Research Centre, School of Mathematics and Physics, Queen's University Belfast, Belfast BT7 1NN, UK}



\begin{abstract}

A thermonuclear explosion triggered by a helium-shell detonation on a carbon-oxygen white dwarf core has been predicted to have strong UV line blanketing at early times due to the iron-group elements produced during helium-shell burning.
We present the photometric and spectroscopic observations of SN~2016dsg, a sub-luminous peculiar Type I SN consistent with a thermonuclear explosion involving a thick He shell. With a redshift of 0.04, the $i$-band peak absolute magnitude is derived to be around -17.5. 
The object is located far away from its host, an early-type galaxy, suggesting it originated from an old stellar population.
The spectra collected after the peak are unusually red, show strong UV line blanketing and weak \ion{O}{1}~$\lambda$7773 absorption lines, and do not evolve significantly over 30 days. 
An absorption line around 9700-10500 \AA{} is detected in the near-infrared spectrum and is likely from the unburnt helium in the ejecta. 
The spectroscopic evolution is consistent with the thermonuclear explosion models for a sub-Chandrasekhar mass white dwarf with a thick helium shell, while the photometric evolution is not well described by existing models.

\end{abstract}



\section{Introduction} \label{sec:intro}
Thermonuclear explosions involving white dwarfs (WDs) have often been associated with Type Ia supernovae (SNe Ia) \citep{Whelan1973, Iben1984,Webbink1984, Nomoto1984, Branch1995}. One of the promising channels to trigger such an explosion is through the detonation of a helium layer atop a sub-Chandrasekhar/near-Chandrasekhar mass WD. In some cases, the initial helium-shell (He-shell) detonation triggers the detonation of the carbon-oxygen (CO) WD core, dubbed double detonation \citep{Nomoto1982_he_shell, Livne1990,Livne1991,Woosley1986,Woosley1994,Livne1995,Hoeflich1996,Nugent1997}.
Historically, the observational features produced by a double detonation with a thick He shell are thought to be quite different from what we see in normal Type Ia SNe \citep{Woosley1994,Hoeflich1996,Nugent1997}. However, recent theoretical work suggests that a double detonation could lead to a normal SN Ia if the He shell is thin enough and, in some studies, polluted by carbon from the underlying CO WD core \citep{Fink2010, Kromer2010,Woosley2011,Shen2014_double,Townsley2019, Polin2019, Gronow2020, Boos2021, Magee2021,Shen2021}. By varying the mass of the He shell and the WD, double detonations can lead to a variety of observational properties and have been used to explain different peculiar sub-types of SNe Ia, including OGLE-2013-SN-079 \citep{Inserra2015}, SN~2016jhr \citep{Jiang2017}, SN~2018byg \citep{De2019}, SN~2016hnk \citep[although see \citealt{Galbany2019}]{Jacobson2020} and SN~2019yvq
\citep[although see \citealt{Miller2020} and \citealt{Burke2021}]{Siebert2020}.

A detonation may only happen in the He shell and fail to trigger a subsequent WD core detonation; this scenario has been dubbed a single detonation or a He-shell detonation \citep{Nomoto1982_he_shell, Woosley1986, Bildsten2007, Shen2010, Waldman2011, Sim2012, Dessart2015}. While it has been shown that the secondary CO WD core detonation is likely to be robustly triggered by even a thin He shell \citep{Fink2010,Shen2014_o/ne_core}, a pure He-shell detonation still could happen for a low-mass CO WD or a O/Ne WD \citep{Shen2014_o/ne_core}.
A pure He-shell detonation produces a faint and fast SN and has been invoked to explain calcium-strong transients \citep{Perets2010,Waldman2011} and a handful of other peculiar transients \citep{Poznanski2010,Kasliwal2010, Inserra2015}. To avoid confusion, we will use `single detonation' to refer to a pure He-shell detonation, in which the initial detonation of the He shell does not trigger the secondary detonation of the CO core.



In this paper, we present the light curves and spectra of SN~2016dsg, a sub-luminous peculiar Type I SN consistent with a thermonuclear explosion involving a thick He shell. The paper is organized as follows: the observations of SN~2016dsg are presented in Section \ref{sec:observations}. We compare SN~2016dsg with single/double detonation models in Section \ref{sec:model_comparison}. In Section \ref{sec:discussion}, we discuss the implications of the observational data, and finally we present our conclusions in Section \ref{sec:conclusions}.

\begin{figure*}
\includegraphics[width=1.\linewidth]{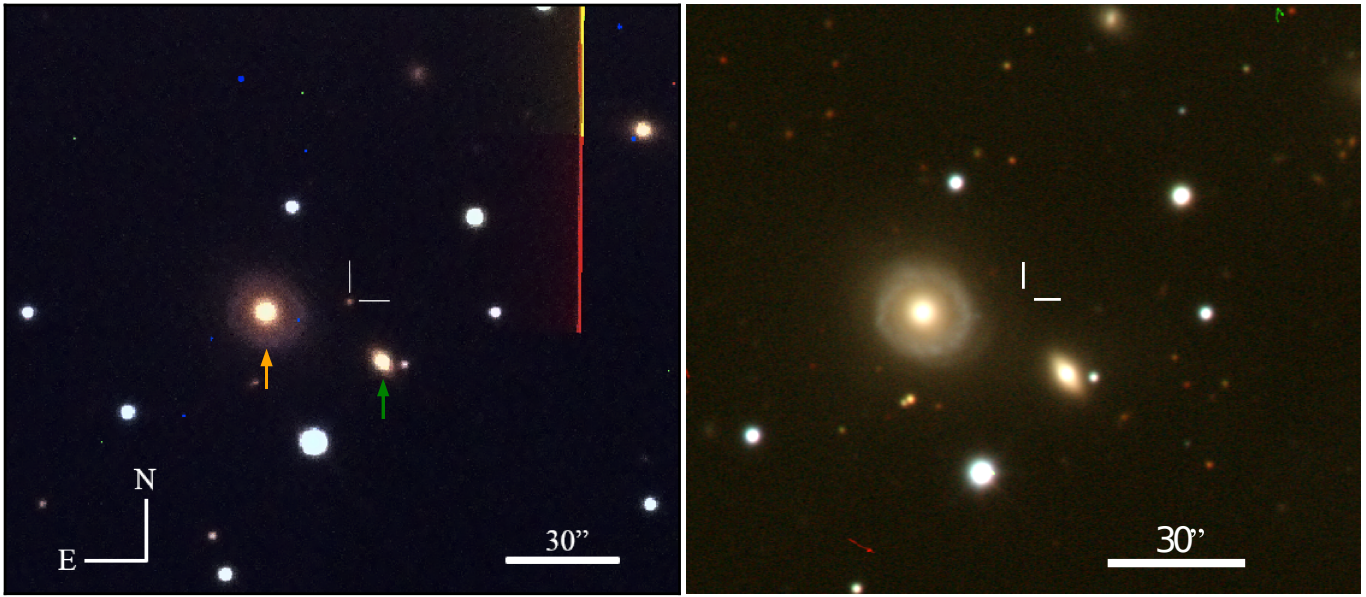}
\caption{Left: Composite $BVr$ image of SN~2016dsg obtained with the Las Cumbres Observatory on 2016 February 27. The two possible host galaxies, ESO 254- G 019 (orange arrow) and WISEA J060707.31-451108.4 (green arrow), are indicated. 
Right: Archival image of the field of SN~2016dsg from the DESI Legacy Imaging Surveys.
The position of SN~2016dsg is indicated by white tick marks in both images.
\label{fig:sn_image}}
\end{figure*}

\begin{deluxetable*}{ccccccc}
\tablecaption{Spectroscopic observations of SN~2016dsg\label{tab:spectra}}
\tablewidth{0pt}
\tablehead{
\colhead{UT Date} & \colhead{Julian Date (Days)} & \colhead{Phase (Days)} &
\colhead{Telescope} & \colhead{Instrument} & \colhead{Resolution ($\lambda/\Delta\lambda$)} & wavelength range (\AA)}
\startdata 
2016-02-26&2457444.66&4.7&NTT&EFOSC2&355&3639-9232\\
2016-02-27&2457445.97&6.0&FTS&FLOYDS&400-700&5000-9000\\
2016-03-02&2457450.38&10.4&SALT&RSS&360&3695-9196\\
2016-03-07&2457454.66&14.7&NTT&EFOSC2&355&3637-9231\\
2016-03-09&2457456.60&16.6&NTT&SofI&550&9377-16468\\
2016-03-14&2457462.33&22.3&SALT&RSS&360&3696-9197\\
2016-03-19&2457467.33&27.3&SALT&RSS&360&3697-9197\\
2016-04-10&2457489.27&49.3&SALT&RSS&360&3798-8196\\
2016-04-15&2457494.25&54.2&SALT&RSS&360&3798-8196\\
\enddata{}
\tablecomments{Phase with respect to the discovery date}
\end{deluxetable*}

\section{Observations} \label{sec:observations}
SN~2016dsg was discovered by the {\it Gaia} Photometric Alert System \citep{Hodgkin2021} on 2016 February 21 (JD~2\,457\,440.02) at RA(2000) $=$ 06$^h$07$^m$08$^s$.18, DEC(2000) = $-$45\degr 10\arcmin 52\farcs 23 as Gaia16afe and was given the IAU name SN~2016dsg \citep{Delgado2016TNSTR.481....1D}. 
SN~2016dsg will be used to refer to the source hereafter. 
Two nearby galaxies, ESO~254-~G~019 (z = 0.040, \citealt{Jones2004, Jones2009}) and WISEA~J060707.31-451108.4 (z = 0.039, \citealt{Jones2004, Jones2009}), are identified within 1 arcmin radius of the SN~2016dsg.
SN~2016dsg is at a projected offset of 22\farcs 49 (18.5 kpc) and 18\farcs 75 (15.4 kpc) from these two galaxies, respectively  (see Figure \ref{fig:sn_image}). 
The Milky Way line-of-sight reddening toward SN~2016dsg is $E(B-V)$ = 0.099 mag \citep{Schlafly2011}.
Given that the object is far from both plausible host galaxies and there are no obvious narrow Na I D absorption lines from the optical spectra \citep{Poznanski2012}, we assume there is no host galaxy extinction.
Throughout the paper, we will adopt a redshift of z = 0.04, corresponding to a luminosity distance of 169 Mpc with $\rm H_0$ = 73 km~$\rm s^{-1}$~$\rm Mpc^{-1}$, $\Omega_{\rm M}$ = 0.27 and $\Omega_{\Lambda}$ = 0.73 \citep{Spergel2007}.

\begin{figure}
\includegraphics[width=1\linewidth]{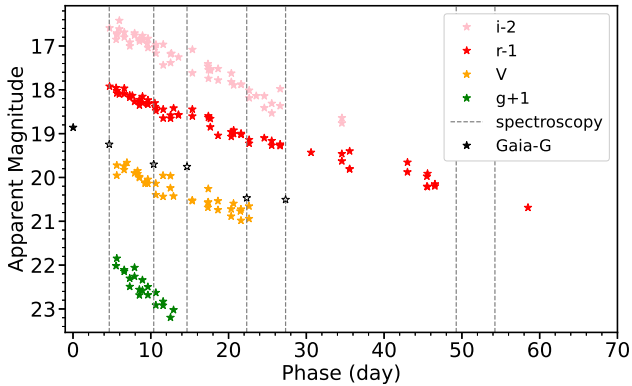}
\caption{Photometric evolution of SN~2016dsg with respect to the epoch of discovery. The phase is measured from the discovery. The vertical lines indicate the epochs when the spectroscopy was taken. For the Gaia $G$-band light curve, the first data point is from direct Gaia observation and the rest of them (hollow stars) are calculated by performing synthetic photometry on spectra. \label{fig:photo}}
\end{figure}

The follow-up observations started from 2016 February 26. Photometric data were obtained through Global Supernova Project with the Las Cumbres Observatory \citep{Brown13} and were reduced with the PyRAF-based photometric reduction pipeline {\sc lcogtsnpipe} \citep{Valenti2016}. The final PSF magnitudes were calibrated using the APASS \citep{Henden2012} catalog stars in the SN field. The background contamination was removed with {\sc HOTPANTS} \citep{Becker2015} by subtracting reference images obtained on 2020 August 23, over four years after the explosion. All the photometric observations are listed in Table \ref{tab:photometry_data}.

Two optical spectra and one near-infrared (NIR) spectrum were obtained by the PESSTO collaboration \citep{smartt2015A&A...579A..40S}. The optical spectra were taken with the ESO Faint Object Spectrograph and Camera (v.2) \citep[EFOSC2;][]{Buzzoni1984} on the 3.6-m New Technology Telescope (NTT). The NIR spectrum was taken with the Son of ISAAC infrared spectrograph and imaging camera \citep[SofI;][]{Moorwood1998} on the NTT. All these spectra were reduced using the PESSTO pipeline as described in \cite{smartt2015A&A...579A..40S}. In addition, five optical spectra were collected with the Southern African Large Telescope (SALT) using the Robert Stobie Spectrograph \citep[RSS;][]{Smith2006} with a 1$\farcs$5 longslit. We used the PG0300 grating in two tilt angles to cover the optical wavelength range and a custom data reduction pipeline based on the PySALT package \citep{Crawford2010}. 
One low-dispersion optical spectrum was obtained on 2016-02-27 by the FLOYDS spectrograph \citep{Brown13}. 
The FLOYDS spectrum was reduced following standard procedures using the {\sc FLOYDS} pipeline \citep{Valenti2014_floyds}.
However, this spectrum had a low signal-to-noise ratio, so we did not use it for analysis. The optical spectra were calibrated to interpolated r-band photometry, and the resulting flux uncertainty is about 10 percent.
All the spectroscopic observations are listed in Table \ref{tab:spectra} and will be available on WISeREP \citep{Yaron2012}\footnote{\url{http://www.weizmann.ac.il}}.

\begin{figure}
\includegraphics[width=1\linewidth]{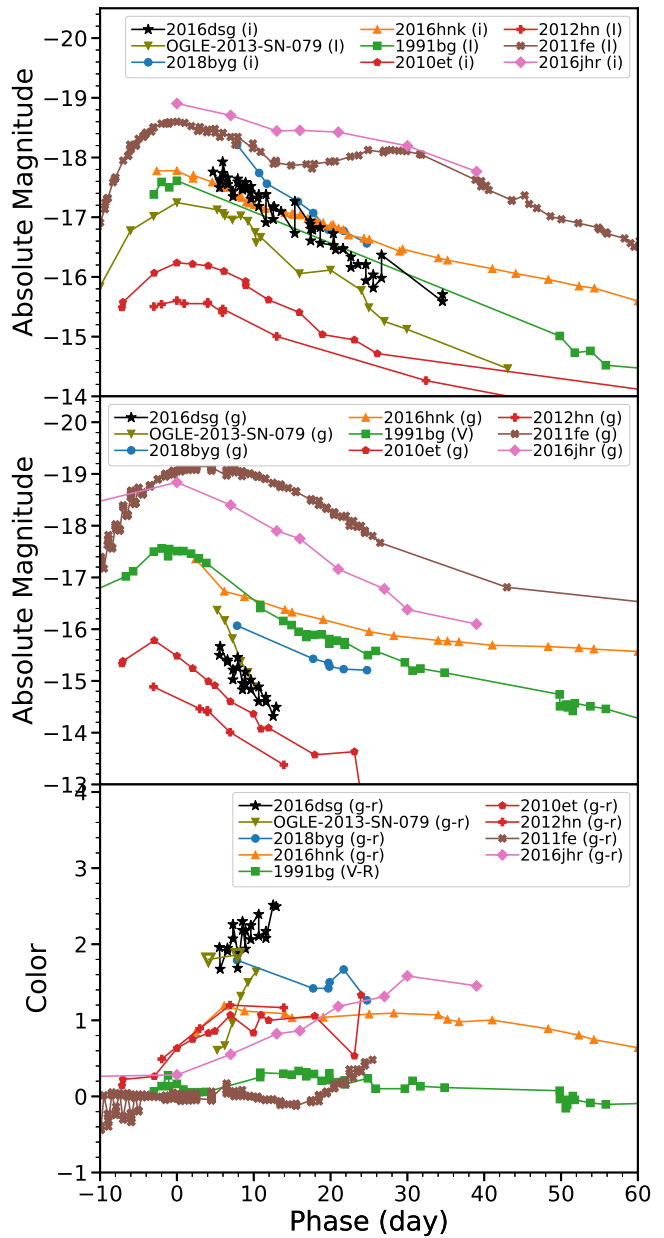}
\caption{Top: $i/I$-band comparison of SN~2016dsg and various subtypes of thermonuclear SNe. SN~2016dsg is fainter than normal SNe Ia but brighter than Ca-strong transients. Middle: $g/V$-band comparison of SN~2016dsg and various subtypes of thermonuclear SNe. Bottom: Color comparison of SN~2016dsg and various subtypes of thermonuclear SNe. The hollow triangles are derived from synthetic photometry of OGLE-2013-SN-079. Around the peak light, SN~2016dsg, OGLE-2013-SN-079 and SN~2018byg are redder than all other objects in our sample. All the objects are extinction corrected. The Vega magnitudes have been converted to AB magnitude system.
}\label{fig:photo_obs_com}
\end{figure}

\begin{figure*}
\includegraphics[width=1\linewidth]{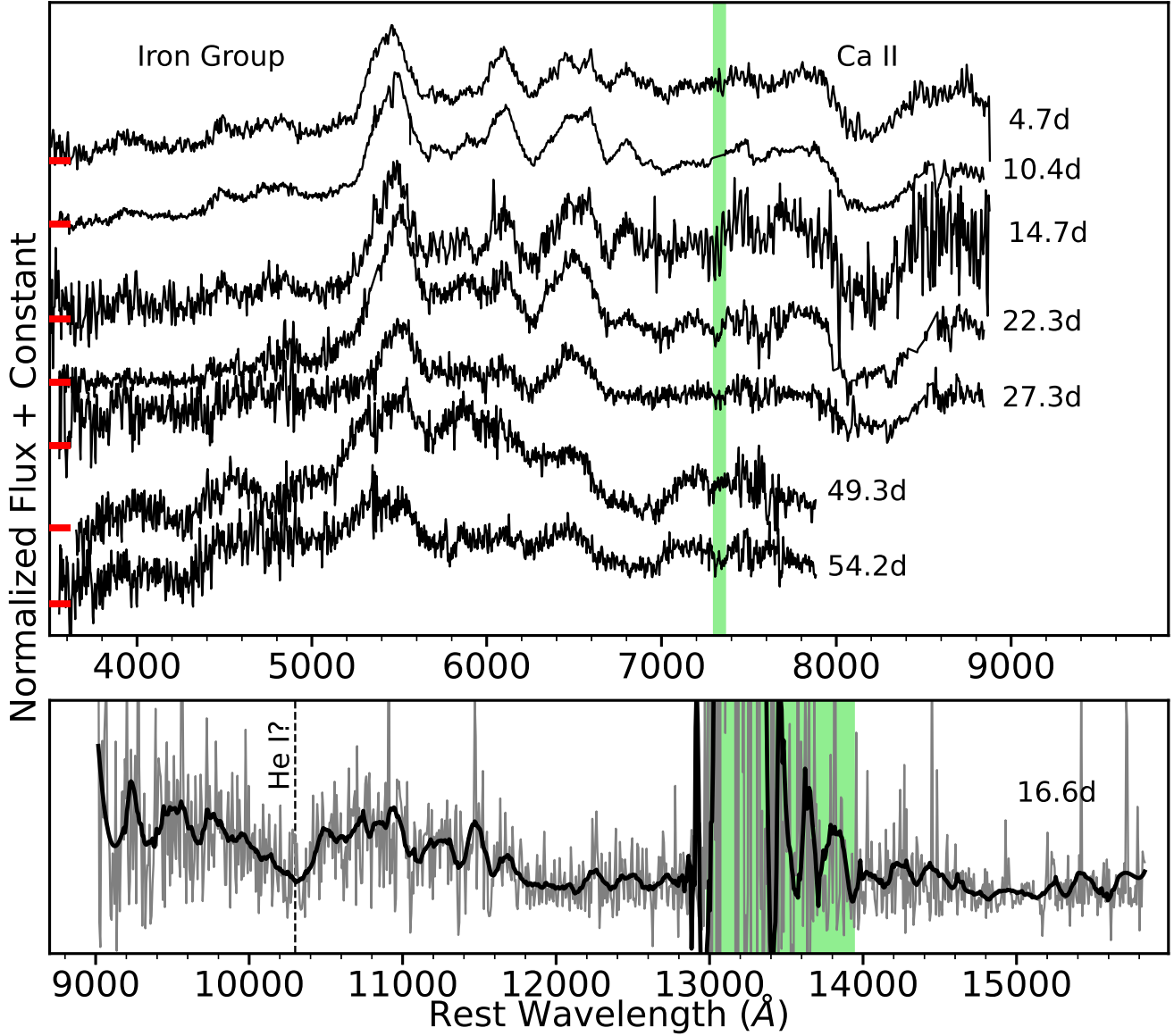}
\caption{Top: spectroscopic evolution of SN~2016dsg. The zero flux levels of each spectrum are indicated by the red ticks. All the spectra have been normalized in the range 4000-7000\AA. Bottom: The NIR spectrum taken at day 16.6. The spectrum has been smoothed with a second order Savitzky-Golay filter, and the gray background line is the original spectrum. The green band marks the strongest telluric absorptions.}
\label{fig:spec}
\end{figure*}

\begin{figure*}[t]
\includegraphics[width=0.98\linewidth]{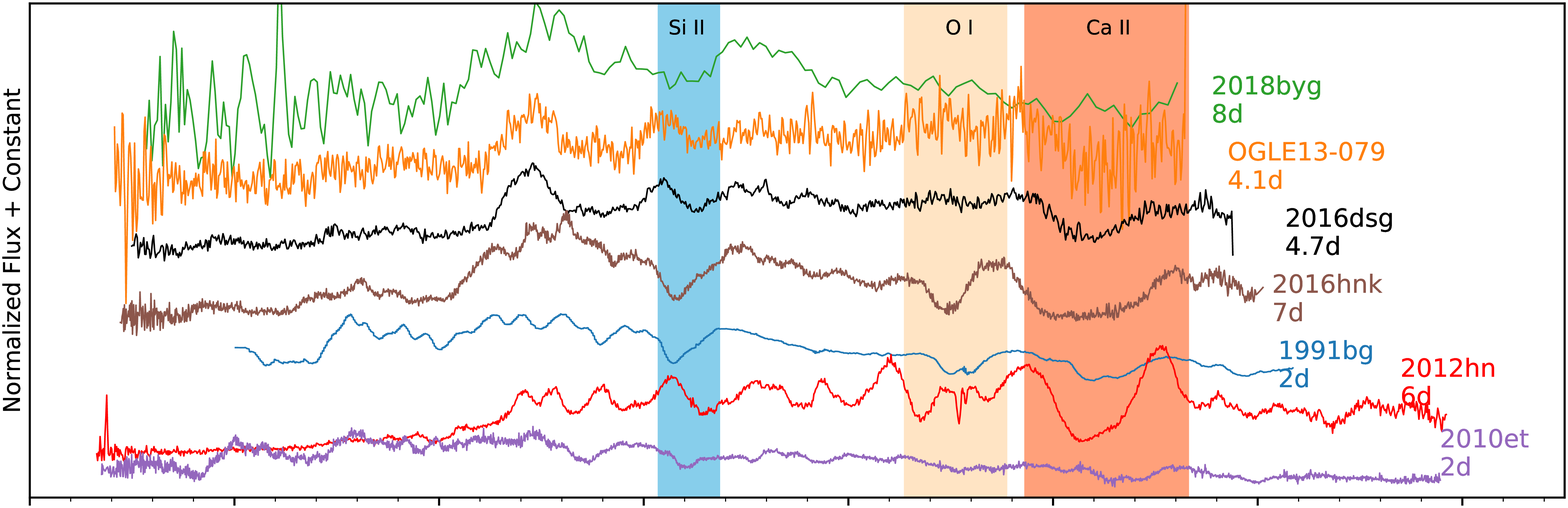}
\includegraphics[width=0.98\linewidth]{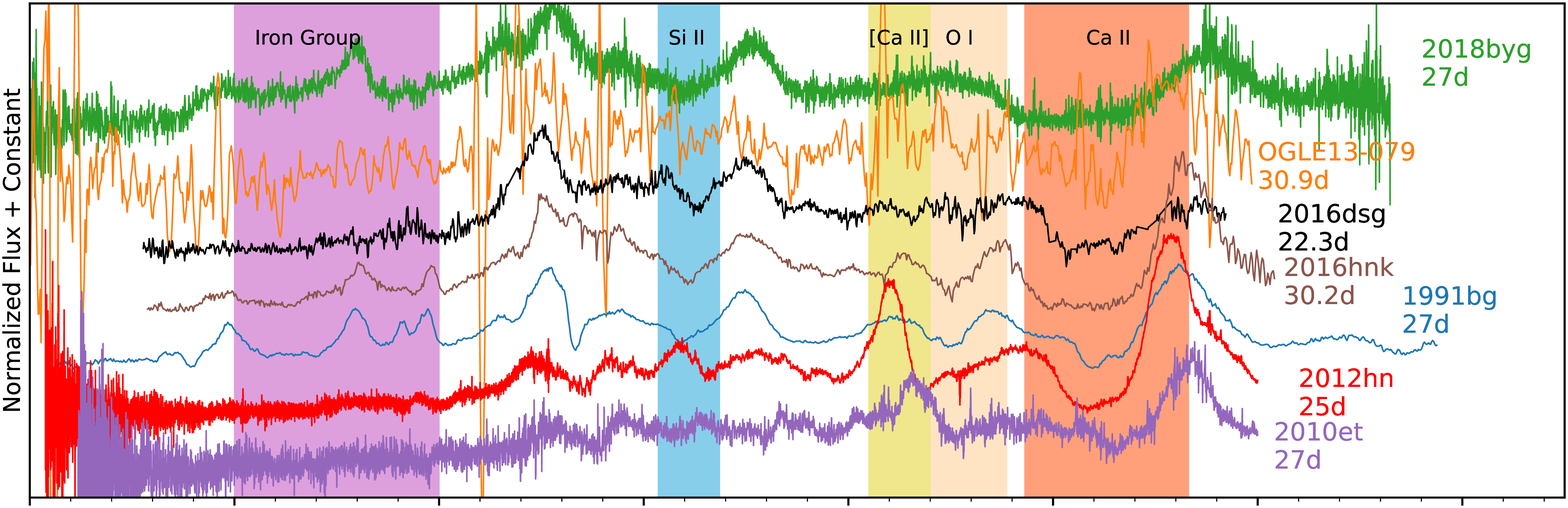}
\includegraphics[width=0.98\linewidth]{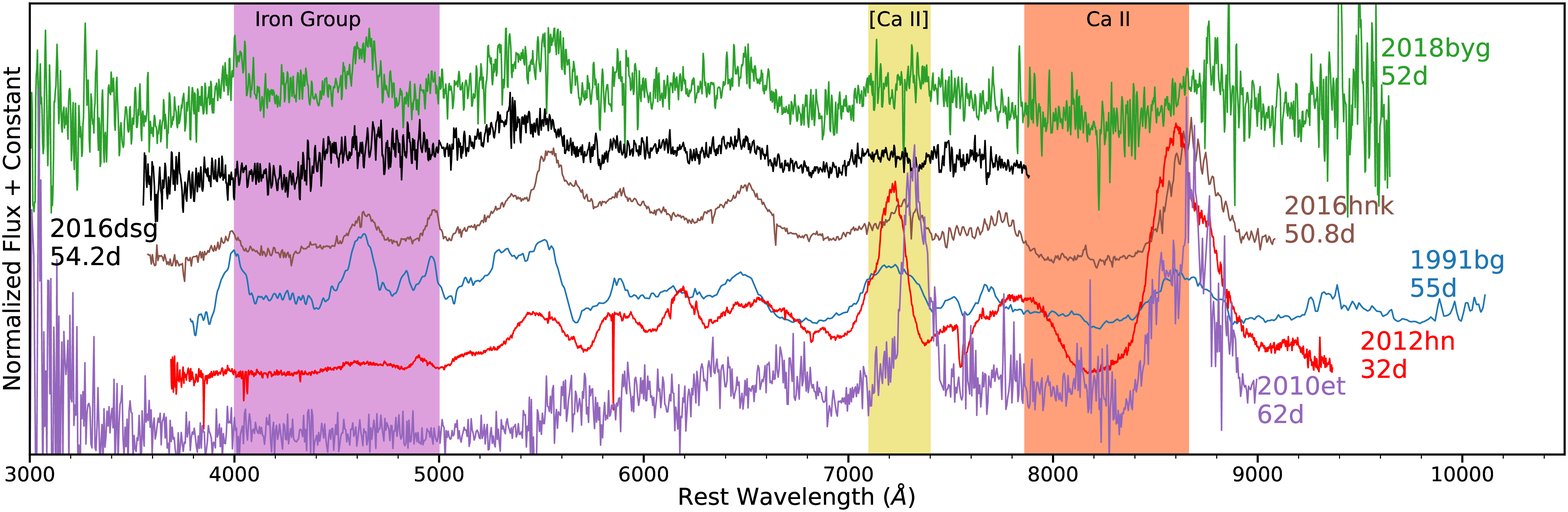}
\caption{Optical spectral comparison of SN~2016dsg to other thermonuclear SNe, including 1991bg-like SNe Ia: SN~1991bg and SN~2016hnk; Ca-strong transients: SN~2012hn and SN~2010et; single/double detonation candidate: OGLE-2013-SN-079 and double detonation candidate: SN~2018byg.}\label{fig:spec_obs_com}
\end{figure*}

\begin{figure}
\includegraphics[width=1\linewidth]{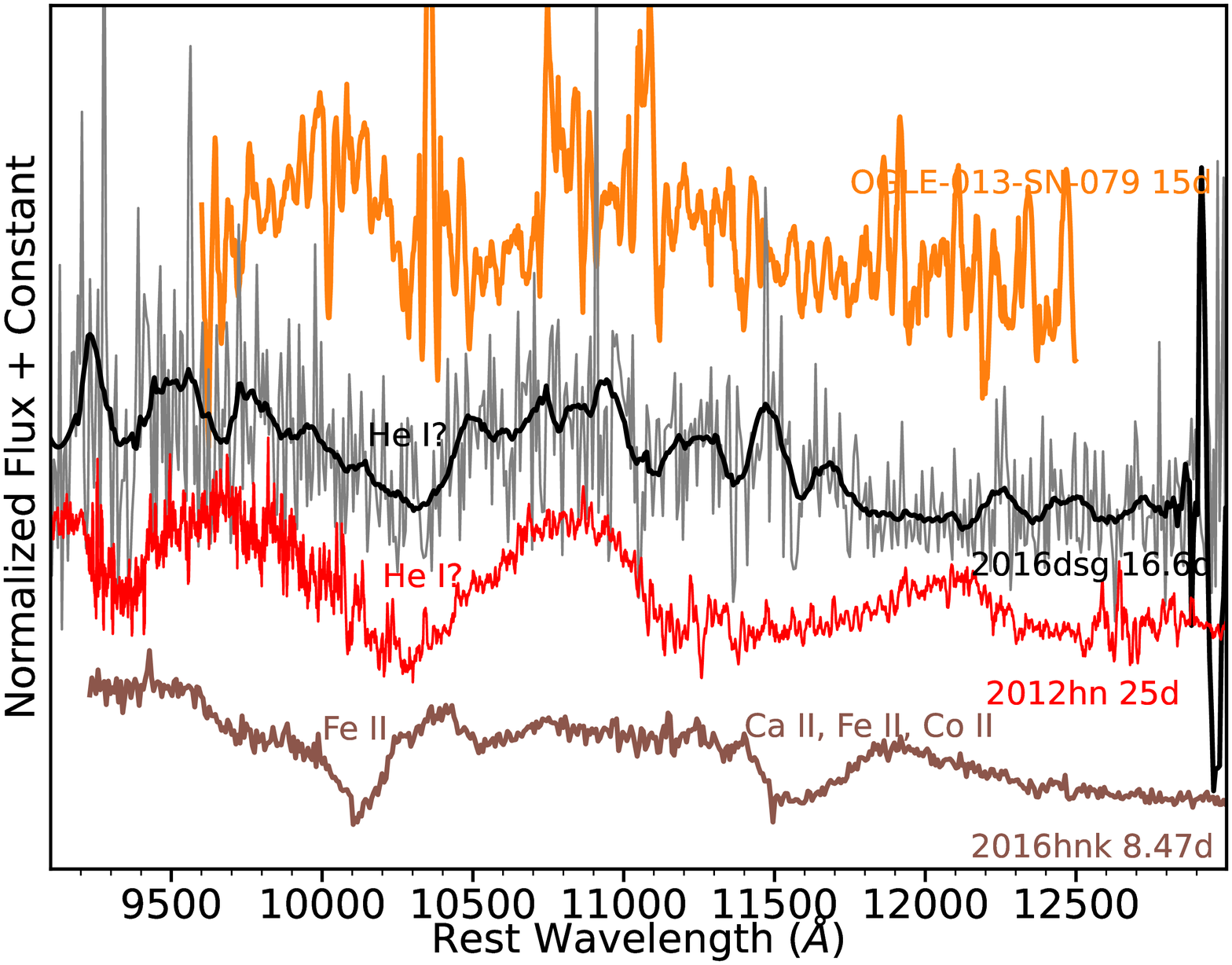}
\caption{The NIR spectrum of SN~2016dsg compared to other transients. A He I absorption feature is likely present in SN~2016dsg.}\label{fig:nir_com}
\end{figure}

\subsection {Photometric Evolution} \label{sec:photometry}
The light curves of SN~2016dsg are shown in Figure \ref{fig:photo}. 
The photometry we collected does not show the rise phase, and the decline is almost linear. 
In the top panel of Figure \ref{fig:photo_obs_com}, we compare the $i$-band light curve of SN~2016dsg with $i/I$-band light curves of other thermonuclear SNe with various subtypes: normal Type Ia SN~2011fe \citep{Nugent2011}; 1991bg-like SN Ia: SN~1991bg \citep{Filippenko1992}; Ca-strong 1991bg-like SN Ia: SN~2016hnk \citep{Galbany2019}; Ca-strong transients: SN~2012hn \citep{Valenti2014} and SN~2010et \citep{Kasliwal2012}; single/double detonation candidate: OGLE-2013-SN-079 \citep{Inserra2015} and double detonation candidates: SN~2018byg \citep{De2019} and SN~2016jhr \citep{Jiang2017}. 
The phase is measured from the discovery for SN~2016dsg and from the i/I-band maximum for other objects. 
The $g/V$-band light curve comparison and $g-r$/$V-R$ color comparison are shown in the middle and bottom panel of Figure \ref{fig:photo_obs_com}, respectively. In order to compare photometric data in AB and Vega system, the V- and I-band magnitudes have been shifted to the AB magnitude system by applying a zero point correction.
In i/I band, SN~2016dsg is less luminous than normal SNe Ia and SN~2016jhr, brighter than some Ca-strong transients and comparable to 1991bg-like transients (SN~1991bg and SN~2016hnk), OGLE-2013-SN-079 and SN~2018byg. The $i$-band decline rate of SN~2016dsg is around 0.077(0.003) mag/day, similar to those of single/double detonation candidates and slightly faster than other objects.
In g/V-band, SN~2016dsg shows a faster decline rate than other objects except for OGLE-2013-SN-079.
At $\sim$ day 4.7, the $g-r$ color of SN~2016dsg is $\sim$1.8, redder than normal SNe Ia, 1991bg-like transients and Ca-strong transients. It is likely that the g-band photometry of OGLE-2013-SN-079 at first a few epochs was affected by a zero point issue, leading to a bluer color. Therefore, we derived the synthetic photometry from spectroscopy, and the resulting $g-r$ colors are similar to those of SN~2016dsg and SN~2018byg at similar epochs.

As shown in Figure~\ref{fig:photo}, SN~2016dsg declined continuously in the Gaia unfiltered white light $G$ bandpass ($G$-band, \citealt{Jordi2010}) based on synthetic photometry derived from our spectra after initial detection, implying that maximum light occurred before discovery. However, as will be discussed in Section \ref{sec:connections}, the epoch of discovery of SN~2016dsg, i.e., JD~2\,457\,440.0, is likely close to the epoch of maximum light, which we adopt as the reference epoch throughout the paper. In this case, the maximum $i$-band absolute magnitude of SN~2016dsg is about $-$17.5 mag, fainter than normal SNe Ia.

\subsection {Spectroscopic Evolution} \label{sec:spectroscopy}
The optical spectra are shown in the upper panel of Figure \ref{fig:spec}. 
These spectra are all collected after the peak brightness, and the spectral features do not evolve too much. The most prominent feature is the strong line blanketing on the blue side of the spectra. Such a feature has been observed in the spectra of OGLE-2013-SN-079 and SN~2018byg, and has been attributed to iron-group elements \citep{Inserra2015, De2019}. 
In the bottom panel of Figure \ref{fig:spec}, we show the NIR spectrum taken at day 16.6, where the spectrum is smoothed with a Savitzky-Golay filter \citep{Savitzky1964}. 

In Figure \ref{fig:spec_obs_com}, we compare the optical spectra of SN~2016dsg at day 4.7, day 22.3 and day 54.2 with those of other thermonuclear SNe at similar epochs. OGLE-2013-SN-079 and SN~2016dsg have almost identical spectroscopic features, implying that these two objects may have very similar progenitors. 
SN~2018byg also has strong UV line blanketing and shows slightly higher velocities than SN~2016dsg. Starting from at least day 27, strong emission lines of Fe-group elements appear on the blue side of the spectra of SN~2018byg \citep{De2019}. 
For SN~2016dsg and OGLE-2013-SN-079, such features are not observed. 
This is consistent with what we see in Figure \ref{fig:photo_obs_com}, where SN~2018byg shows a shallower decline in the $g$ band than SN~2016dsg and OGLE-2013-SN-079.
Compared to 1991bg-like transients, SN~2016dsg shows rather weak \ion{O}{1}~$\lambda$7773 absorption. This could be due to the low abundance of oxygen in the ejecta and will be discussed in detail in Section \ref{sec:discussion}.
Ca-strong transients quickly develop strong [\ion{Ca}{2}]~$\lambda\lambda$7291, 7323 emission lines, which are either very weak or do not exist in SN~2016dsg.

In Figure \ref{fig:nir_com}, we show a NIR spectroscopic comparison between SN~2016dsg, SN~2012hn, OGLE-2013-SN-079 and SN~2016hnk. 
SN~2016dsg has a broad absorption line at 9700-10500 \AA, which is likely to be \ion{He}{1}~$\lambda$10830.
SN~2012hn and SN~2016hnk have similar features in this region, but this line has been identified as Fe II for SN~2016hnk \citep{Galbany2019}. 
This possible \ion{He}{1}~$\lambda$10830 absorption feature in SN~2016dsg will be discussed in Section \ref{unburned_he}.
 




\begin{deluxetable*}{cccccccc}
\tablecaption{A brief summary of the models which were compared to SN~2016dsg. \label{tab:model}}
\tablewidth{0pt}
\tablehead{
\colhead{} \vspace{-0.2cm} &
\colhead{Original} & 
\colhead{Detonation} & \colhead{WD} &
\colhead{He Shell} &
\colhead{Total} &
\colhead{Nickel} &
\colhead{} \\ 
\colhead{Model} \vspace{-0.2cm} &
\colhead{Model} & 
\colhead{Type} & \colhead{Mass} &
\colhead{Mass} &
\colhead{Mass } &
\colhead{Mass } &
\colhead{Reference\tablenotemark{$\star$}} \\
\colhead{} &
\colhead{Name} & \colhead{} & \colhead{(\msun{})} &
\colhead{(\msun{})} &
\colhead{(\msun{})} &
\colhead{(\msun{})} &
\colhead{}
}
\startdata 
Polin0.8+0.08-D& &Double&0.8&0.08&0.88&0.074&a\\
Polin0.9+0.08-D& &Double&0.9&0.08&0.98&0.31&a\\
Polin0.76+0.15-D& &Double&0.76&0.15&0.91&0.18&a\\
Polin0.76+0.15-0.2-D$\dagger$&   &Double &0.76&0.15&0.91&0.18&a\\
Kromer0.81+0.126-D&Kromer model 1 & Double& 0.81 & 0.126 & 0.936&0.17&b\\
Kromer1.08+0.055-D&Kromer model 3 & Double & 1.08 & 0.055 & 1.135&0.55&b\\
Sim0.45+0.21-D&Sim CSDD-L & Double & 0.45 & 0.21 & 0.66&0.0218&c\\
Sim0.58+0.21-D&Sim CSDD-S & Double & 0.58 & 0.21 & 0.79&0.215&c\\
Sim0.58+0.21-S&Sim HeD-S & Single & 0.58 & 0.21 & 0.79&0.065&c\\
Shen0.6+0.2-S&Shen 0.6+0.2 & Single & 0.6 & 0.2 & 0.8&0.026&d\\
Shen1.0+0.10-S&Shen 1.0+0.10 & Single & 1.0 & 0.1 & 1.1&0.0504&d\\
\enddata
\tablenotetext{\star}{(a) \cite{Polin2019}; (b) \cite{Kromer2010}; (c) \cite{Sim2012}; (d) \cite{Shen2010}.}
\tablenotetext{\dagger}{\ Polin0.76+0.15-0.2-D is similar to Polin0.76+0.15-D but includes 0.2 \msun{} mixing in the outer ejecta.}
\end{deluxetable*}
\vspace{-0pt}

\section{Model Comparison}
\label{sec:model_comparison}
Since SN~2016dsg is fainter than normal Type Ia SNe, the possibility that its progenitor is a sub-Chandrasekhar mass WD can not be directly excluded \citep{Sim2010,Blondin2017,Shen2018_subWD}. However, a quick comparison with the models from \cite{Blondin2017} shows that the sub-Chandrasekhar mass models fail to match the level of absorption or line-blanketing below $\sim$5000\AA{} of the observations. The models also predict significant \ion{O}{1}~$\lambda$7773 absorption lines that are not visible or rather weak in SN~2016dsg.
The strong line-blanketing in the UV around the peak light is usually an indication of a large amount of iron-group elements in the outer ejecta and likely points to a single/double detonation. 
In the following section, we will compare the object with single detonation models from \cite{Shen2010} and \cite{Sim2012}, and double detonation models from \cite{Kromer2010}, \cite{Sim2012}, and \cite{Polin2019}. 


In \cite{Shen2010}, the observables of single detonations for various CO WD core masses (0.6, 1.0 and 1.2~\msun) and He envelope masses (0.05, 0.1, 0.2, 0.3~\msun) were explored. 
\cite{Kromer2010} explored the observable properties of double detonation models with minimum mass helium shells (from 0.0035 to 0.0126~\msun) studied in \cite{Fink2010}.
\cite{Sim2012} modeled a low mass system (0.45~\msun{} WD + 0.21~\msun{} He) and a high mass system (0.58~\msun{} WD + 0.21~\msun{} He) for both single detonation and double detonation scenarios. \cite{Polin2019} exploded double detonation models for a set of CO WD masses (from 0.6 to 1.2~\msun) with helium shells of 0.01, 0.05 and 0.08~\msun.
In order to roughly match the brightness and the UV line-blanketing of SN~2016dsg, the models we chose to compare with SN~2016dsg are Polin0.8+0.08-D, Polin0.9+0.08-D, Polin0.76+0.15-D, Polin0.76+0.15-0.2-D, Kromer0.81+0.126-D, Kromer1.08+0.055-D, Sim0.45+0.21-D, Sim0.58+0.21-S, Sim0.58+0.21-D, Shen0.6+0.2-S and Shen1.0+0.10-S. 
In the literature, Sim0.45+0.21-D, Sim0.58+0.21-S and Shen0.6+0.2-S have been found to provide a reasonable match to the spectra of OGLE-013-SN-079 \citep{Inserra2015}. Polin0.76+0.15-0.2-D was compared to SN~2018byg \citep{De2019} and well-reproduced the observations.
We urge the reader to refer to the original references for further details. The original names and parameters of these models are listed in Table \ref{tab:model}.


\begin{figure*}[t]
\includegraphics[width=1\linewidth]{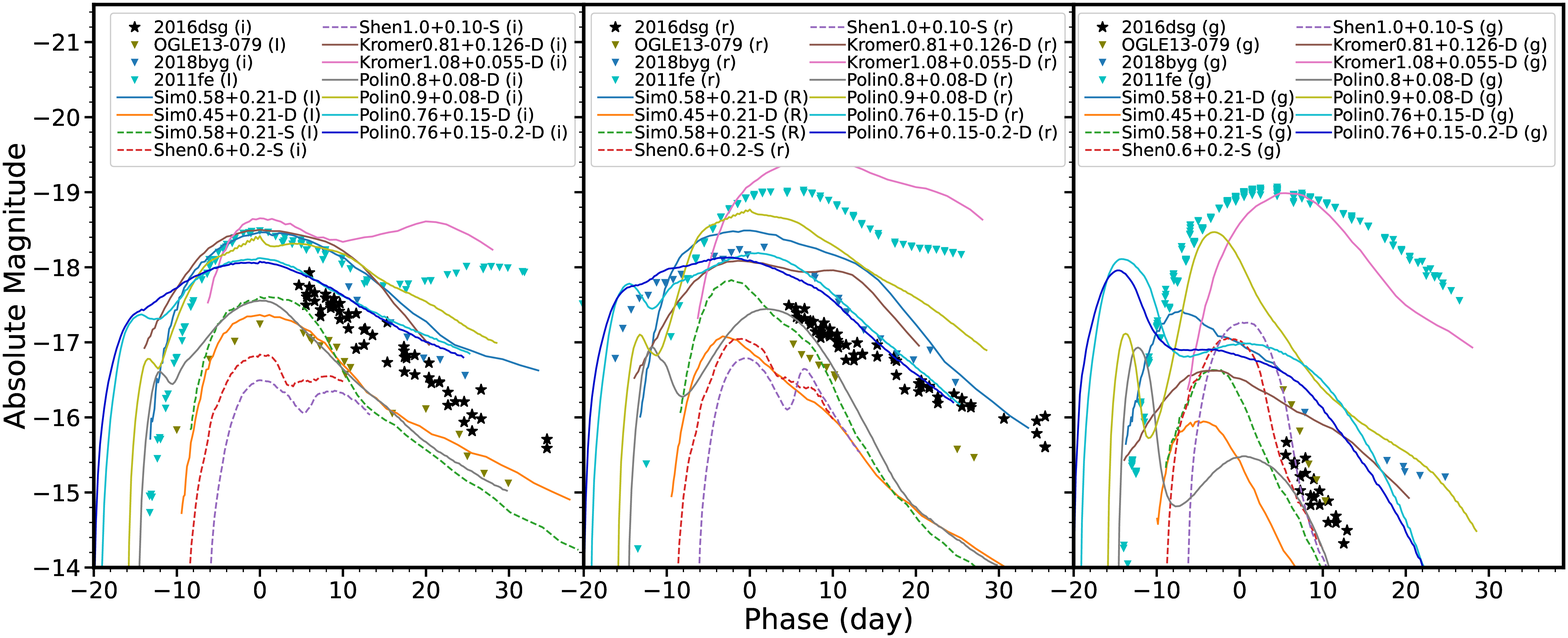}
\includegraphics[width=1\linewidth]{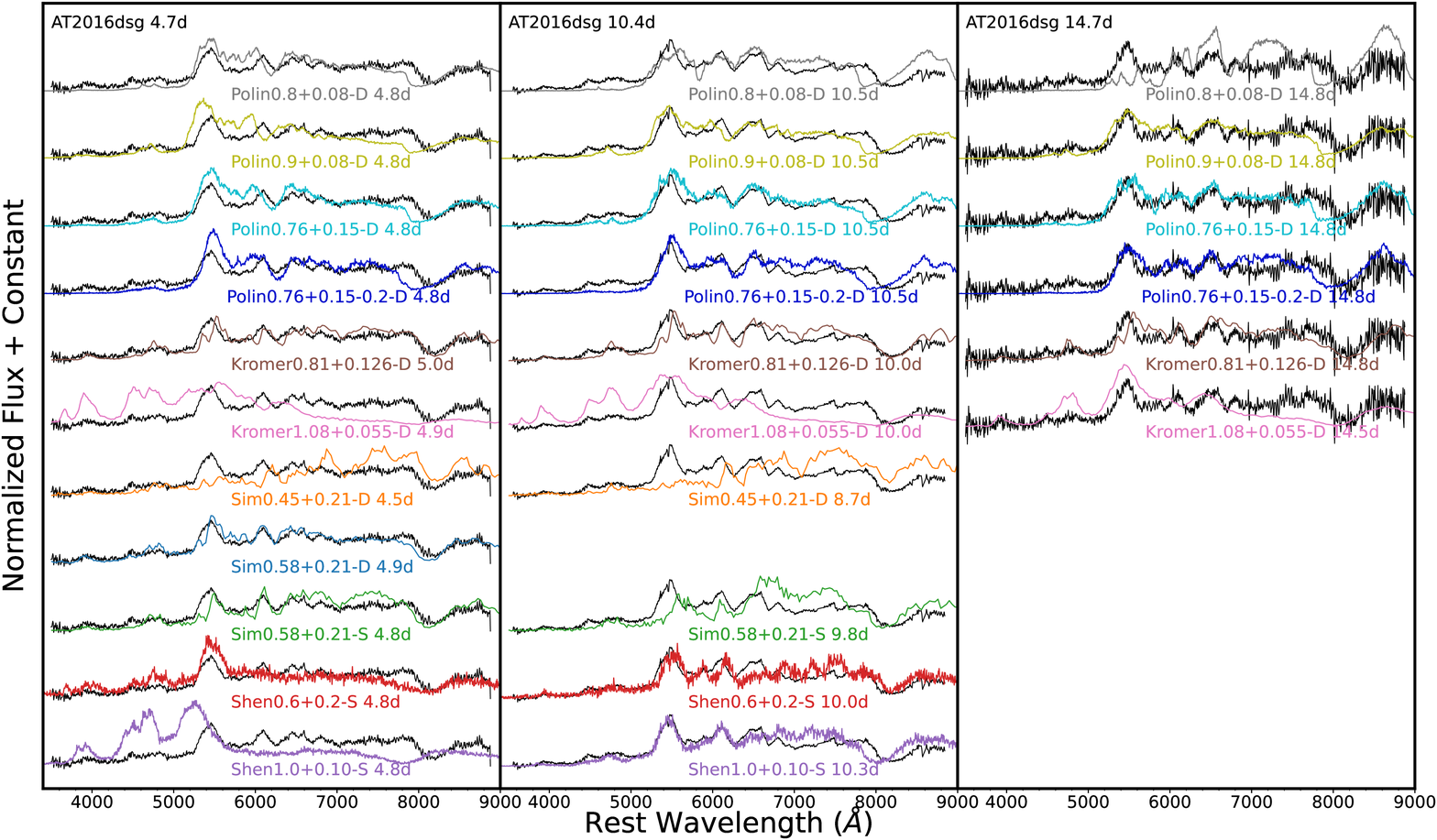}
\caption{Top: light curve comparison between SN~2016dsg, other transients and models. 
The single detonation models and double detonation models are plotted with dashed lines and solid lines, respectively. 
The phase is measured from the $i/I$-band maximum.
Bottom: Spectroscopic comparison between SN~2016dsg and single/double detonation models. All the spectra have been normalized in the range 4000-8500\AA. \label{fig:model_com}}
\end{figure*}

\subsection{Light Curves}
Since the detonation of a He shell is usually not symmetric, the viewing angle of the observer could have a significant influence on the observed light curves \citep{Kromer2010, Sim2012,Gronow2020, Shen2021}. The effect of viewing angle gets stronger for bluer bands and lower-mass progenitors \citep{Shen2021}; for this reason we focus on comparisons in the $i/I$ band, which is less affected. In the top left panel of Figure \ref{fig:model_com}, we compare the $i$-band light curve of SN~2016dsg with $i/I$-band light curves of models and other transients. We also show the $r/R$-band and $g$-band comparison in the top right panel of Figure \ref{fig:model_com} for the reference. 

All the single detonation model light curves in the plot (Sim0.58+0.21-S, Shen0.6+0.2-S and Shen1.0+0.10-S) are slightly fainter than the observed light curves of SN~2016dsg. 
For the $i/I$ band, Polin0.76+0.15-D and Polin0.76+0.15-0.2-D roughly reproduce the brightness and the slope of the light curve. Sim0.45+0.21-D and Polin0.8+0.08-D underestimate the brightness of the light curve, while Polin0.9+0.08-D, Kromer0.81+0.126-D, Kromer1.08+0.055-D and Sim0.58+0.21-D overestimate the brightness. 
For the $r/R$ band, Kromer0.81+0.126-D, Kromer1.08+0.055-D, Sim0.58+0.21-D, Polin0.9+0.08-D, Polin0.76+0.15-D and Polin0.76+0.15-0.2-D overestimate the brightness but reproduce the slope of SN~2016dsg. The rest of the models fail to reproduce both the brightness and the slope of the observed light curve.

In general, none of these models match the observed light curves well. However, the shape of the light curves is influenced by many factors, such as the mass of WD core and He shell \citep{Shen2010,Kromer2010,Sim2012,Polin2019}, the mixing degree of the outer layers \citep{Polin2019, Gronow2020} and the viewing angle \citep{Kromer2010, Sim2012,Gronow2020, Shen2021}. In addition, the models we are comparing to are all from Local-Thermodynamic-Equilibrium (LTE) simulations or simplified non-LTE simulations, an assumption that is not suitable at very late time when the ejecta are optically thin. In order to nicely reproduce the whole light curves, a non-LTE simulation and a fine tuning to the WD and He-shell masses and other parameters would be needed.

\subsection{Spectra}
We compare the spectra of SN~2016dsg with angle-averaged model spectra in the bottom panel of Figure \ref{fig:model_com}. 
As we described in Section \ref{sec:spectroscopy}, the spectra of SN~2016dsg show very strong line-blanketing on the blue side, which is reproduced by most of the models except for Kromer1.08+0.055-D. 
The Polin0.9+0.08-D, Polin0.76+0.15-D and Polin0.76+0.15-0.2-D models roughly fit the spectra of SN~2016dsg, while the Polin0.8+0.08-D model does not reproduce the spectral feature at day 14.7. All of Polin's models show higher calcium velocities than SN~2016dsg.
Kromer0.81+0.126-D well reproduces the line strengths and velocities of SN~2016dsg.
Kromer1.08+0.055-D shows low level absorption in the UV, which is likely because this model has a rather thin He shell (0.055 \msun) and the initial detonation of the He shell does not produce enough iron-group elements to absorb photons in the UV.
Sim0.45+0.21-D does not reproduce features around 5000-7000 \AA. Sim0.58+0.21-D and Sim0.58+0.21-S are consistent with the observations at day 4.7, but Sim0.58+0.21-S slightly overpredicts the flux at around 6500-7500 \AA. 
Shen0.6+0.2-S is able to reproduce most of the line features at day 4.7 and day 10.4, but is overall too blue relative to SN~2016dsg at day 4.7. Shen1.0+0.10-S shows low level absorption below $\sim$5000 \AA{} at $\sim$day 4.8, while it well reproduces the observational spectra at day 10.3.
Despite some small discrepancies, most of the models generally reproduce the main features of SN~2016dsg. This supports the idea that a He shell is likely to be involved for the progenitor of SN~2016dsg.

\subsection{Progenitor Implications}
The brightness of the object is roughly proportional to the total mass of the progenitor for double detonations or the He-shell mass of the progenitor for single detonations.
However, there are some differences among the different models that we should take into account. For example, for a similar total mass, Sim's models are always brighter than others.
This is likely due to the more complete burning in their models \citep{Sim2012}. 
In addition, \cite{Sim2012} assumed that the He shell is composed of pure He in their models. In reality, the He layer can be polluted by the WD core material \citep{Piro2015}, which would reduce the amount of radioactive material produced by He burning \citep{Kromer2010,Waldman2011,Townsley2019, Gronow2020, Magee2021} and thus lead to a fainter transient at early times. Therefore, for a double detonation, Sim's models give the brightest event for a certain total progenitor mass. 
This can be used to constrain the lower limit of the progenitor's total mass in the double detonation scenario. The Sim0.45+0.21-D model is slightly fainter than SN~2016dsg in $i/I$ band, so any double detonation models that have a total mass less than Sim0.45+0.21-D would be fainter than SN~2016dsg. As a result, we obtain a total mass lower limit of $\sim$0.7\msun{} for SN~2016dsg. 
On the other hand, SN~2016dsg is fainter than Polin0.9+0.08-D, Polin0.76+0.15-D, Polin0.76+0.15-0.2-D, Kromer0.81+0.126-D and Kromer1.08+0.055-D, implying that the total mass of the progenitor is no more than $\sim$0.9\msun.
Similarly, for a single detonation, Sim's model will also be the brightest for a given He-shell mass. The Sim0.58+0.21-S  model roughly matches the brightness of SN~2016dsg, so the He-shell mass of SN~2016dsg should be no less than $\sim$0.2\msun{} for a single detonation.

The early time light curves of single/double detonations are powered by the radioactive material produced by the He-shell, so the early time photometric and spectroscopic data can be used to estimate the He-shell mass. In addition, the He-shell mass of the progenitor could also be roughly constrained by a spectral comparison with models. We found that when the He-shell mass is less than $\sim$0.1\msun{} (Polin0.8+0.08-D , Kromer1.08+0.055-D and Shen1.0+0.10-S), the model cannot match the observed spectra well, suggesting that the He-shell mass of the progenitor should be larger than $\sim$0.1\msun{} for either a double or a single detonation. However, we note that, without the early time photometric and spectroscopic data, the He-shell mass derived here is just a rough estimation.

It has been suggested that even a very low-mass He shell can detonate and then trigger the following WD detonation \citep{Bildsten2007,Fink2010}, inevitably leading to a double detonation, while \cite{Waldman2011} argued that the 
robustness of the second detonation in a double detonation scenario may need further investigations. \cite{Shen2014_o/ne_core} showed that a system with a low-mass CO WD or a O/Ne WD core is harder to be ignited by the initial He-shell detonation, so theoretically a single detonation can exist in nature. For SN~2016dsg, we are not able to distinguish between the single detonation scenario and the double detonation scenario, so both of them could be used to explain this object.

Due to the radioactive decay of elements synthesized in the He shell, the double detonation light curve may show an early red excess during the first few days \citep{Noebauer2017,Maeda2018,Polin2019}. This phenomenon has been observed in SN~2016jhr \citep{Jiang2017} and could be a unique signature for the double detonation scenario. Nevertheless, the lack of early photometric data prevents us from extracting more information from the photometry.

\begin{figure*}[t]
\includegraphics[width=1.\linewidth]{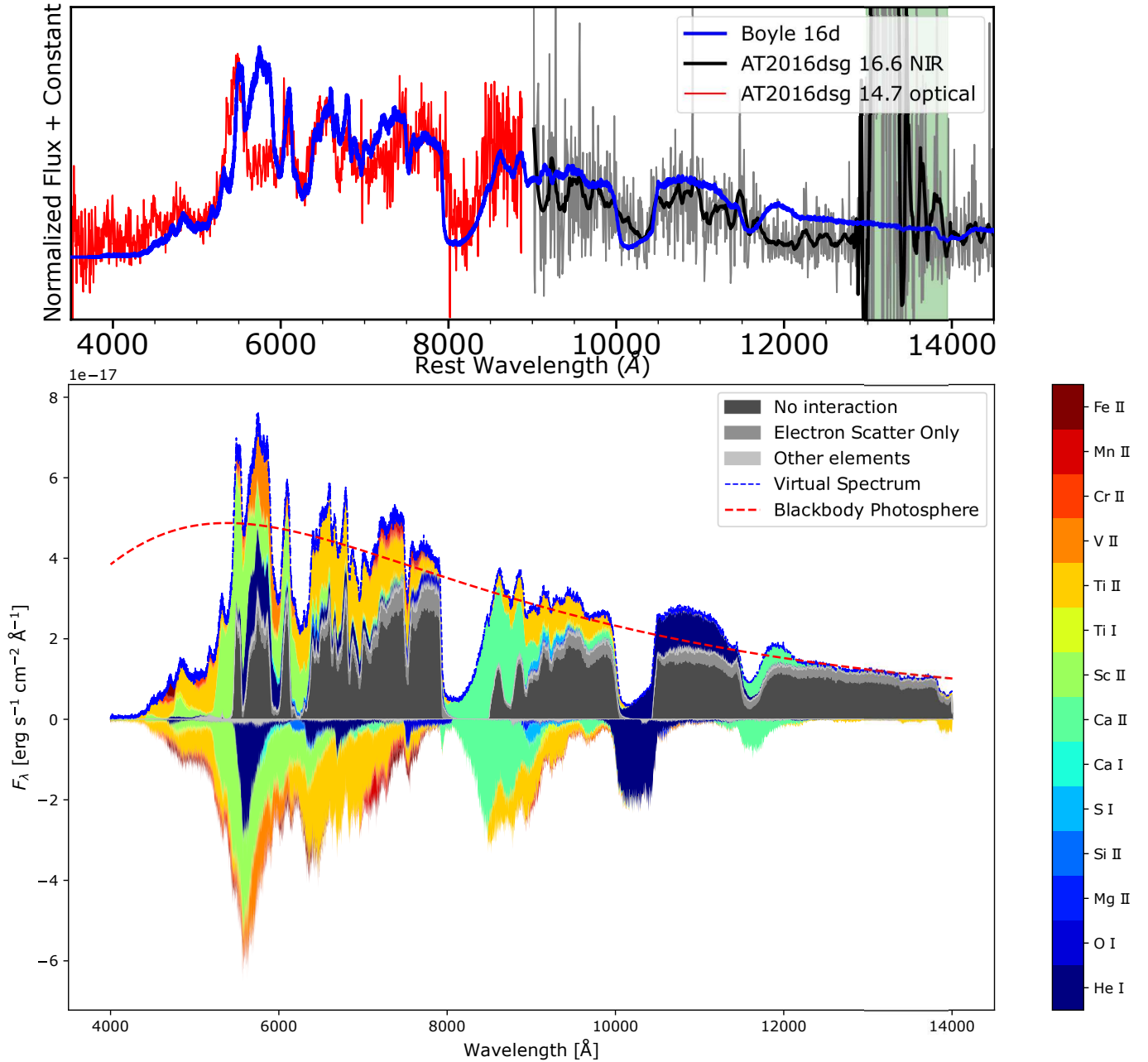}
\caption{Top: Comparison of the optical-NIR spectrum of SN~2016dsg to the low mass double detonation model from \cite{Boyle2017}. The green band marks the strongest telluric absorptions. Bottom: The SDEC plot produced by {\sc TARDIS}. Different colors are used to illustrate the contributions from different elements. Absorption is represented by negative values. ``Interaction'' in this context refers to radiation-matter interaction above the blackbody photosphere; ``no interaction'' means that photons escape freely.} \label{fig:element_decom}
\end{figure*}

In conclusion, the comparison between models and observational data suggests that the progenitor of SN~2016dsg should have a total mass of 0.7-0.9 \msun{} with a He-shell mass larger than 0.1 \msun{} for a double detonation and a He-shell mass larger than 0.2 \msun{} for a single detonation.

\section{Discussion} \label{sec:discussion}

\subsection{Unburnt Helium?} \label{unburned_he}

Recent simulations have revealed that unburnt helium exists in the outer ejecta of single detonations as well as double detonations
\citep{Fink2010,Shen2010,Kromer2010,Sim2012,Shen2014_o/ne_core,Polin2019}.
By performing non-LTE simulations, \cite{Dessart2015} found that the spectral lines from unburnt He can be seen in the single detonation, with the most prominent one being the \ion{He}{1}~$\lambda$10830 line. In their model, \ion{He}{1}~$\lambda$10830 has a P-cygni profile and is visible beyond 5 days after the explosion. However, the light curves produced by their model do not fit those of SN~2016dsg. 
Their I-band model light curve peaks at -16.1 mag, almost 2 mag fainter than our object.
The observational signature of unburnt helium for double detonations has been explored by \cite{Boyle2017} using the Monte Carlo radiative transfer code {\sc TARDIS} \citep{Kerzendorf2014}. They predict that a high velocity \ion{He}{1}~$\lambda$10830 line can be detected around maximum light in double detonations. In their work, they used \cite{Kromer2010}'s model 3 (referred to as Kromer1.08+0.055-D in this paper) as their high mass model and \cite{Sim2012}'s CSDD-S (referred to as Sim0.58+0.21-D in this paper) as their low mass model, and the high velocity \ion{He}{1}~$\lambda$10830 line appears much stronger in Sim0.58+0.21-D than in Kromer1.08+0.055-D.

As discussed in Section \ref{sec:model_comparison}, Sim0.58+0.21-D at day 4.7 can reproduce the optical spectra of SN~2016dsg well.
In order to examine whether a strong high velocity He I feature is present in SN~2016dsg, we generated a new model spectrum at 16 days after the i-band peak based on the model setup described in \cite{Boyle2017} with their low mass model (Sim0.58+0.21-D). Following \cite{Boyle2017}, the inner velocity boundary in {\sc TARDIS} is set to be 6250 $\rm km~s^{-1}$.
A comparison between this model spectrum and the observed spectrum is shown in the top panel of Figure \ref{fig:element_decom}. As a sanity check, we also overplotted the optical spectrum at day 14.7. In the bottom panel of Figure \ref{fig:element_decom}, we show a Spectral element DEComposition (SDEC) plot of the {\sc TARDIS} model. In the SDEC plot, the contributions of different elements in the synthetic model spectrum are illustrated by different colors. Negative values correspond to absorption contributions to the spectrum, while positive values indicate emission.
The absorption line we see at around 9700-10500 \AA{} in SN~2016dsg resembles the strong He I line 
in the model, implying that there is likely unburnt He present in the ejecta. This is consistent with our previous conclusion that SN~2016dsg could originate from a double detonation.

\cite{Boyle2017} pointed out that the \ion{He}{1}~$\lambda$10830 line can be used to distinguish between double detonations and single detonations. In double detonations, the unburnt He all resides in the outer ejecta and is only at high velocities, while in single detonations, the unburnt He is distributed at all velocities. Therefore, the emission part of the \ion{He}{1}~$\lambda$10830 line is shallower and broader in double detonations than in single detonations. However, due to the lack of a comparable single detonation model and the low signal-to-noise ratio of the observed spectrum, we are not able to use the NIR spectrum to distinguish between the single detonation scenario and the double detonation scenario.

\subsection{Connections to Other single/Double Detonation Candidates?} 
\label{sec:connections}
\begin{figure}
\includegraphics[width=1\linewidth]{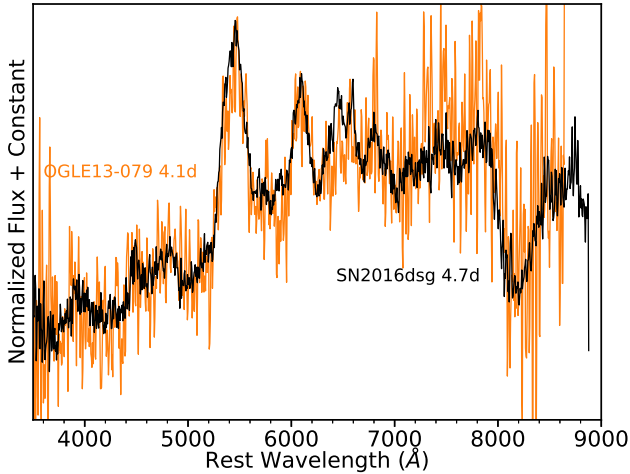}
\caption{The spectrum of SN~2016dsg compared to OGLE-2013-SN-079. These two objects have similar spectroscopic features, suggesting that they have similar progenitors. \label{fig:ogle_com}}
\end{figure}

OGLE-2013-SN-079 and SN~2018byg are thought to originate from single/double detonations with thick He shells \citep{Inserra2015, De2019}, and there are indeed many similarities between these two objects and SN~2016dsg.
The spectrum of OGLE-2013-SN-079 is remarkably similar to the spectrum of SN~2016dsg (see Figure \ref{fig:ogle_com}), implying that these two objects have very similar progenitors.
SN~2018byg developed strong emission lines below 5000 \AA{} after at least day 27, which is not observed in the spectra of SN~2016dsg or OGLE-2013-SN-079. 
This discrepancy could be solved by taking the viewing angle and the He-shell mass of the progenitor into account.
Many authors have shown that the viewing angle has a significant impact on the spectra of double detonations \citep{Kromer2010, Sim2012, Gronow2020, Shen2021}. Specifically, the spectrum will be redder if one observes from the direction of the He-shell ignition point, where the He-shell burning is more efficient and thus produces more iron-group elements. It is possible that SN~2016dsg and OGLE-2013-SN-079 were observed closer to the ignition point than SN~2018byg, resulting in the higher level absorption below 5000 \AA.
In addition, a larger mass He shell is able to produce more iron-group elements, blocking the emission in the UV. Therefore, it is also possible that SN~2016dsg and OGLE-2013-SN-079 have more surface He than SN~2018byg does.
On the other hand, SN~2018byg had higher expansion velocities than SN~2016dsg. If SN~2018byg were a double detonation, the He-shell ashes would likely become optically thin earlier than SN~2016dsg, and could develop emissions at relatively earlier phases.

Due to the similarities between SN~2016dsg and OGLE-2013-SN-079, it is natural to expect that these two objects also have similar photometric evolution. In Figure \ref{fig:photo_obs_com}, SN~2016dsg's phase is shown with respect to the epoch of discovery, and it approximately matches the evolution of OGLE-2013-SN-079. If the maximum epoch is much earlier than the discovery date, SN~2016dsg would be much brighter than OGLE-2013-SN-079. Therefore, the epoch of discovery of SN~2016dsg is likely close to the true epoch of maximum light.

The origin of SN~2016hnk is still debated. \cite{Jacobson2020} proposed that SN~2016hnk is from a double detonation with a relatively thin He shell (0.02 \msun), while \cite{Galbany2019} found that this object is consistent with a near Chandrasekhar-mass CO WD explosion. 
Comparing the spectra, SN~2016hnk has a much stronger \ion{O}{1}~$\lambda$7773 absorption line than SN~2016dsg, OGLE-2013-SN-079 and SN~2018byg. 
The weak O I line is an indication of a lack of oxygen in the outer ejecta and could be a natural result of a single or double detonation with a thick He shell. 
For a single detonation, only a negligible amount of oxygen is produced \citep{Shen2010, Waldman2011, Sim2012, Dessart2015}, so the ejecta have little oxygen. For a double detonation, the outermost ejecta are the ash of He-shell burning products and the unburnt oxygen from the core is confined underneath with a narrow velocity range \citep{Kromer2010, Sim2012, Hoeflich2017, Polin2019}. If the He shell is massive enough, the unburnt oxygen would reside in a deep layer of the ejecta and thus may not produce an obvious spectroscopic feature at early phases.
Therefore, the strong O I line in SN~2016hnk may imply that either SN~2016hnk is not from a double detonation, or that it is from a double detonation with a very low-mass He shell. If SN~2016hnk is from a double detonation, the diversity observed between SN~2016hnk and the other three objects could be achieved by varying the thickness of the He shell.

Compared to other sub-luminous SNe Ia, Ca-strong transients develop strong [Ca II] lines shortly after the peak. 
One promising model for Ca-strong transients is the single detonation \citep{Perets2010, Waldman2011,Dessart2015,Shen2019}. 
In addition, \cite{Polin2021} found that double detonations, analogous to Ca-strong transients, are also expected to show strong [Ca II] emission at the nebular phase.
A large proportion of Ca-strong transients have large offsets from their host galaxies \citep{Perets2010,Kasliwal2012,Valenti2014,De2020}, which is similar to the other three promising single/double detonation candidates (see Table \ref{tab:host}). This may suggest that there is a connection between the progenitors of Ca-strong transients and SN~2016dsg.

\begin{deluxetable*}{cccc}
\tablecaption{Summary of host galaxy properties\label{tab:host}}
\tablewidth{0pt}
\tablehead{
\colhead{Object} & \colhead{Host} & \colhead{Host type} &
\colhead{projection distance from the host}}
\startdata 
SN~2016dsg&ESO254-G019/WISEA J060707.31-451108.4&S0(r)&15.4/18.5kpc\\
SN~2018byg&-&elliptical&21.9kpc\\
OGLE-013-SN-079&2MASXJ00350521-6741147&elliptical&~40-50kpc\\
\enddata{}
\tablecomments{Host galaxy properties of SN~2016dsg and other two spectroscopically similar transients.}
\end{deluxetable*}

\subsection{Remote Location}
SN~2016dsg shares many similarities with OGLE-2013-SN-079 and SN~2018byg. Interestingly, they are all in the outskirts of their host galaxies, suggesting an origin in old stellar populations. 
Large host galaxy offsets have also been observed in many Ca-strong transients, and have been treated as evidence that the progenitors of Ca-strong transients could have travelled a long distance before exploding \citep{Lyman2014,Lyman2016}.
Although a sample of three SNe precludes any robust statistical analysis, the offsets of these objects are very typical of Ca-strong transients' offsets ($\sim$10-100kpc) \citep{De2020}.
\cite{Foley2015} proposed that the progenitors of Ca-strong transients are double WD systems that are ejected from the hosts through the interactions with supermassive black holes.
\cite{Shen2019} found that the projected galactocentric radial distribution of Ca-strong transients is consistent with that of globular clusters. Therefore, they suggested that the progenitors of Ca-strong transients are formed inside globular clusters and ejected prior to mass transfer contact.
The remote location of SN~2016dsg may also be explained by the mechanisms mentioned above. 

Two hot subdwarf B binaries with white dwarf companions have been suggested as possible progenitors of double/single detonations \citep{Geier2013,Kupfer2022}. 
These systems are found in young populations, which is inconsistent with the fact that the three thick He-shell double/single detonation candidates are found in old populations. If sdB+WD binaries are the progenitors of these double/single detonations, they must have traveled a long way before detonating. However, the sample of double/single detonations and their possible sdB binary progenitors is still small, preventing further investigations at this point.

\section{Conclusions}
\label{sec:conclusions}
We have presented spectroscopic and photometric observations of SN~2016dsg, a sub-luminous peculiar SN I consistent with many of the hallmarks of the single/double detonation explosion mechanism, particularly those models with a thick He shell.  Spectroscopic observations below $\sim$5000 \AA{} show significant line blanketing, indicative of a large amount of iron-group elements in the outer ejecta. 
Compared to other sub-luminous SNe Ia, SN~2016dsg shows a rather weak \ion{O}{1}~$\lambda$7773 absorption, likely due to a low oxygen abundance in the ejecta. 
In addition, a absorption line around 9700-10500 \AA{} is detected in the NIR spectrum. We argue this absorption line is from the unburnt He present in the outer ejecta.
All of these unique observational features suggest that SN~2016dsg came from a single/double detonation.

We have compared SN~2016dsg with many single/double detonation models and found that the spectroscopic properties of SN~2016dsg can be reproduced well by these models. In order to fit the strong line blanketing below $\sim$5000 \AA, a thick He shell has to be involved. For a double detonation scenario, the progenitor should have a total mass of around 0.7-0.9 \msun{} with a He-shell mass of $\gtrsim$ 0.1 \msun. For a single detonation scenario, the mass of the He shell should be no less than 0.1 \msun, and the WD core is very likely to be a low-mass CO WD or an O/Ne WD. 

To date, only a handful of candidate single/double detonation of WDs with thick He shells have been discovered, implying that they may be intrinsically rare in nature. The preference of these objects for remote locations is still a mystery and may give us a clue about the origin of their progenitors. 
The diversity of these transients in the existing sample could be explained by the variety of He-shell masses, WD-core masses, viewing angles, and the degree of mixing in the outer layers, which, however, is to be confirmed by more observational evidence. With the help of upcoming large sky surveys, we can expect that a statistical study of large samples of these types of transients will soon be in reach, which will help us further understand the nature of these peculiar transients.

\begin{acknowledgments}

Based on observations collected at the European Organisation for Astronomical Research in the Southern Hemisphere under ESO programme 1103.D-0328.
This research made use of \textsc{tardis}, a community-developed software package for spectral
synthesis in supernovae \citep{Kerzendorf2014, kerzendorf_wolfgang_2022_5979739}. The
development of \textsc{tardis} received support from GitHub, the Google Summer of Code
initiative, and from ESA's Summer of Code in Space program. \textsc{tardis} is a fiscally
sponsored project of NumFOCUS. \textsc{tardis} makes extensive use of Astropy and Pyne.
This work made use of the Heidelberg Supernova Model Archive (HESMA), https://hesma.h-its.org”.
This research has made use of the NASA/IPAC Extragalactic Database (NED), which is operated by the Jet Propulsion Laboratory, California Institute of Technology, under contract with the National Aeronautics and Space Administration.

Research by Y.D., and S.V., and N.M.R is supported by NSF grants AST–1813176 and AST-2008108. 
Time domain research by D.J.S.\ is also supported by NSF grants AST-1821987, 1813466, 1908972, \& 2108032, and by the Heising-Simons Foundation under grant \#2020-1864.
The SALT observations presented here were taken as part of Rutgers University program 2015-1-MLT-002 (PI: Jha).
This work makes use of observations from the Las Cumbres Observatory network. The Las Cumbres Observatory team is supported by NSF grants AST-1911225 and AST-1911151, and NASA Swift grant 80NSSC19K1639.
L.W. and M.G. acknowledge the Polish National Science Centre (NCN) grants Harmonia No. 2018/30/M/ST9/00311 and Daina No.
2017/27/L/ST9/03221 as well as the European Union's Horizon 2020
research and innovation programme under grant agreement No
101004719 (OPTICON-RadioNet Pilot, ORP) and MNiSW grant
DIR/WK/2018/12.
MG is supported by the EU Horizon 2020 research and innovation programme under grant agreement No 101004719.
L.G. acknowledges financial support from the Spanish Ministerio de Ciencia e Innovaci\'on (MCIN), the Agencia Estatal de Investigaci\'on (AEI) 10.13039/501100011033, and the European Social Fund (ESF) ``Investing in your future" under the 2019 Ram\'on y Cajal program RYC2019-027683-I and the PID2020-115253GA-I00 HOSTFLOWS project, from Centro Superior de Investigaciones Cient\'ificas (CSIC) under the PIE project 20215AT016, and the program Unidad de Excelencia Mar\'ia de Maeztu CEX2020-001058-M.
JDL acknowledges support from a UK Research and Innovation Fellowship (MR/T020784/1).
This work was supported by the `Programme National de Physique Stellaire' (PNPS) of CNRS/INSU co-funded by CEA and CNES.
This research was supported by the Excellence Cluster ORIGINS which is funded by the Deutsche Forschungsgemeinschaft (DFG, German Research Foundation) under Germany's Excellence Strategy EXC-2094-390783311.
S.B. acknowledges support from the ESO Scientific Visitor Programme in Garching.
KM is funded by the EU H2020 ERC grant no. 758638.

\end{acknowledgments}
%

\facilities{Las Cumbres Observatory (Sinistro), 
SAAO:SALT (RSS), 
The New Technology Telescope (EFOSC2, SofI)}


\software{Astropy \citep{astropy13,astropy18}, 
          HOTPANTS \citep{Becker2015},
          lcogtsnpipe \citep{Valenti2016}, 
          Matplotlib \citep{Hunter2007},
          NumPy (https://numpy.org),
          PYRAF,
          Pandas \citep{mckinney-proc-scipy-2010},
          SciPy (https://www.scipy.org),
          TARDIS \citep{Kerzendorf2014,kerzendorf_wolfgang_2022_5979739}
          }



\appendix

\section{Appendix information}
\startlongtable
\begin{deluxetable}{ccccccc}
\tablenum{A1}
\tablecaption{Optical Photometry of SN~2016dsg\label{tab:photometry_data}}
\tablewidth{0pt}
\tablehead{
\colhead{UT Date} & \colhead{Julian Date (Days)} & \colhead{Phase (Days)} &
\colhead{Mag} & \colhead{Mag error} & \colhead{Filter} & \colhead{Source}}
\startdata 
2016-02-26&2457444.70&4.70&18.92&0.07&r&LSC 1m\\
2016-02-26&2457444.70&4.70&18.59&0.08&i&LSC 1m\\
2016-02-27&2457445.54&5.54&21.02&0.19&g&LSC 1m\\
2016-02-27&2457445.54&5.54&18.95&0.17&r&LSC 1m\\
2016-02-27&2457445.55&5.55&18.85&0.15&i&LSC 1m\\
2016-02-27&2457445.63&5.63&19.95&0.16&V&LSC 1m\\
2016-02-27&2457445.63&5.63&19.72&0.17&V&LSC 1m\\
2016-02-27&2457445.63&5.63&20.85&0.22&g&LSC 1m\\
2016-02-27&2457445.64&5.64&19.08&0.09&r&LSC 1m\\
2016-02-27&2457445.64&5.64&19.02&0.07&r&LSC 1m\\
2016-02-27&2457445.64&5.64&18.71&0.14&i&LSC 1m\\
2016-02-27&2457445.64&5.64&18.74&0.14&i&LSC 1m\\
2016-02-27&2457445.95&5.95&19.10&0.14&r&COJ 1m\\
2016-02-27&2457445.96&5.96&18.61&0.14&i&COJ 1m\\
2016-02-27&2457445.96&5.96&18.42&0.13&i&COJ 1m\\
\enddata
\tablecomments{Phase with respect to the discovery date. The full table will be in machine-readable form.}
\end{deluxetable}




\bibliography{gaia16afe}{}
\bibliographystyle{aasjournal}


\end{CJK*}
\end{document}